\documentclass[12pt]{article}
\usepackage{pic03}
\usepackage{graphicx}
\usepackage{cite}

\newcommand   {\e}  {{{e}}}
\newcommand   {\p}  {{{p}}}

\newcommand   {\Z}  {{{Z}}}

\renewcommand {\d}  {{{d}}}

\begin{document}

\title{\bf HARD PROCESSES IN ELECTRON-PROTON SCATTERING}
\author{
Hans-Christian Schultz-Coulon    \\
{\em Institut f\"ur Physik, Universit\"at Dortmund}}
\maketitle

%

\baselineskip=14.5pt

\begin{abstract}
This report summarizes some of the recent HERA results obtained by studying
hard processes in $ep$--scattering.  By resolving the structure of the
proton, hard $ep$--reactions provide information on the parton content of
the proton and may give insight into the dynamics of the exchanged parton
cascade.  In addition, their study offers the possibility to test the
Standard Model, in particular perturbative Quantum Chromodynamics, on which
the theoretical predictions on $ep$--scattering cross sections are generally
based.  Any observed deviation between the data and existing theoretical
models would either indicate the need to calculate higher order
contributions or hint at signs of new physics.
\end{abstract}

\newpage

\baselineskip=17pt

\section{Introduction}

At HERA 27.5~GeV electrons or positrons collide head on with 920~GeV
protons\footnote{The proton beam energy has been increased from 820~GeV to
920~GeV after the 1997 data taking period; data recorded before 1998 are
taken at a center-of-mass energy of 300~GeV.}, leading to a center-of-mass
energy $\sqrt{s}$ of approximately 320~GeV. Due to this large center-of-mass
energy the HERA facility, with its two collider experiments H1 and ZEUS,
offers the possibility to probe the structure of the proton down to very
small distances ($\sim 10^{-18}\;$m).

For a particular $ep$--process the resolving power with which the proton
structure is analyzed, is either given by the virtuality $Q^{2}$ of the
exchanged photon, $Z$-- or $W$--boson, or by any other hard scale,
$\mu^{2}$, inherent to the process studied.  Apart from the investigation of
inclusive deep-inelastic scattering (DIS), studies of hard $ep$--processes
hence also comprise the analysis of exclusive final states, in particular
jet topologies, where substantial resolving power may be provided at
low $Q^{2}$ due to the choice $\mu^{2}=E_{T}^{2}$, with $E_{T}$ representing
the transverse energy of the observed jets.

This report summarizes some of the most recent measurements obtained from
analyzing inclusive DIS data (section~\ref{DIS}) as well as some new HERA results in jet
physics (section~\ref{jets}).  In addition, section~\ref{searches} is dedicated to the search for
new phenomena presenting the unexpected observations of events with a
high-$p_{t}$ isolated lepton and missing transverse momentum and of
multi-electron topologies.

\section{Inclusive Deep-Inelastic Scattering}
\label{DIS}

The Born cross section~\cite{Derman:sp,Ingelman:1987zv} for the neutral
current DIS reaction $\e^{\pm}\p \rightarrow \e^{\pm}X$ is given by
\begin{equation}
\label{eqn=NCgen}
     \frac{\d^2\sigma^{\pm}_{\rm NC}}{\d x\,\d Q^2} =  
           \frac{2\pi\alpha^2}{xQ^4} 
               \left\{ Y_+(y)\,{\cal F}_2(x,Q^2) \mp
                       Y_-(y)\,x{\cal F}_3(x,Q^2) - y^{2}{\cal 
		       F}_{L}(x,Q^2) \right\}
\end{equation}
where $Q^2=-q^2$ is the negative four-momentum transfer squared carried by
the exchanged gauge boson ($\gamma$ or $\Z^0$), $x$ represents the Bjorken
scaling variable, and \mbox{$Y_\pm=1\pm(1-y)^2$} (with $y=Q^2/xs$) describes
the helicity dependence of the electroweak interactions.  The partonic
structure of the proton is then contained in the generalized structure
functions ${\cal F}_2$, ${\cal F}_3$ and ${\cal F}_L$.  For unpolarized
beams ${\cal F}_2$ and ${\cal F}_3$ can be written as
\begin{equation}
\label{eqn=F2def}
  \left(
    \begin{array}{r} 
      {\cal F}_2(x,Q^2) \\ 
      x {\cal F}_3(x,Q^2) 
    \end{array}
  \right) 
  = \sum_{q={\rm quarks}} \!\! x
  \left(
    \begin{array}{c}
      C^q_2(Q^2)\,\left[q(x,Q^2) + \bar{q}(x,Q^2)\right] \\
      C^q_3(Q^2)\,\left[q(x,Q^2) - \bar{q}(x,Q^2)\right]
    \end{array}
  \right) \,\raisebox{-4mm}[-4mm]{.}
\end{equation}
Here, $q$ and $\bar{q}$ are the quark densities depending on $x$ and $Q^2$
alone, while the coefficient functions $C^q_2$ and $C^q_3$ can be expressed
in terms of precisely measured electroweak
parameters~\cite{Ingelman:1987zv}.  The longitudinal structure function
${\cal F}_L$ contributes only at high~$y$ and is related to the gluon
content of the proton.

In contrast to neutral current interactions, for which all quark and
anti-quark flavours contribute, charged current $\e^{-}\p \rightarrow \nu X$
($e^{+}\p \rightarrow \bar{\nu}X$) reactions probe only
up-type (down-type) quarks and down-type (up-type) anti-quarks, as they are
mediated by the exchange of a $W^{-}$ ($W^{+}$) boson.  Charged current
reactions thus allow for flavour-specific investigations of the parton
momentum distributions and can provide additional information on the quark
content of the proton at high~$x$ and high~$Q^{2}$.  Since only the weak
interaction contributes, the expression for the double differential charged
current cross section at the Born-level~\cite{Ingelman:1987zv} can be
written in a somewhat simpler form than in the neutral current case:
\begin{equation}
     \frac{d^2\sigma^{\pm}_{\rm CC}}{dx\,dQ^2} =  
	   \frac{G_{F}^{2} M_{W}^{4}}{2\pi x} \frac{1}{(Q^{2} + 
	   M_{W}^{2})^{2}} \tilde{\sigma}^{\pm}_{\rm CC} 
	   \;\;\raisebox{-1mm}[-1mm]{,}
\end{equation}
with the reduced cross section $\tilde{\sigma}^{\pm}_{\rm CC}$ given by
\begin{eqnarray}
    \tilde{\sigma}^{+}_{\rm CC} & = & x 
       \left[ (\bar{u}(x,Q^{2}) + \bar{c}(x,Q^{2})) + (1-y)^{2} 
       (d(x,Q^{2}) + s(x,Q^{2})) \right] \\
       \tilde{\sigma}^{-}_{\rm CC} & = & x
	  \left[ (u(x,Q^{2}) + c(x,Q^{2})) + (1-y)^{2} (\bar{d}(x,Q^{2}) + 
	  \bar{s}(x,Q^{2})) \right] \;\; \raisebox{-1.5mm}[-1.5mm]{.}
\end{eqnarray}
Here $u$, $c$, $d$ and $s$ are the quark and $\bar{u}$, $\bar{c}$, $\bar{d}$
and $\bar{s}$ the anti-quark distributions.  From these equations the
sensitivity of charged current reactions to the different quark densities
becomes evident; at high~$x$, where the contribution from the sea can be
neglected, $\e^{-}\p$ scattering provides direct access to the $u$-quark
distributions, while positron-proton collisions probe the $d$-quark content
of the proton.

The neutral current (NC) and charged current (CC) cross sections have
been measured by both the H1~\cite{Adloff:2000qk,Adloff:2000qj,%
Adloff:2003uh,H1_EPS01_lowQ2,H1_EPS03_lowQ2} and the ZEUS~\cite{%
Chekanov:2001qu,Chekanov:2002ej,Chekanov:2002zs,Chekanov:2003vw}
collaborations.  From these measurements information on the proton structure
is derived in the form of structure functions and parton distributions.

\subsection{$F_{2}$--Measurements and Parton Distributions}

\begin{figure}[tb]
\begin{center}
  \begin{minipage}{0.6\textwidth}
  \includegraphics[width=\textwidth]{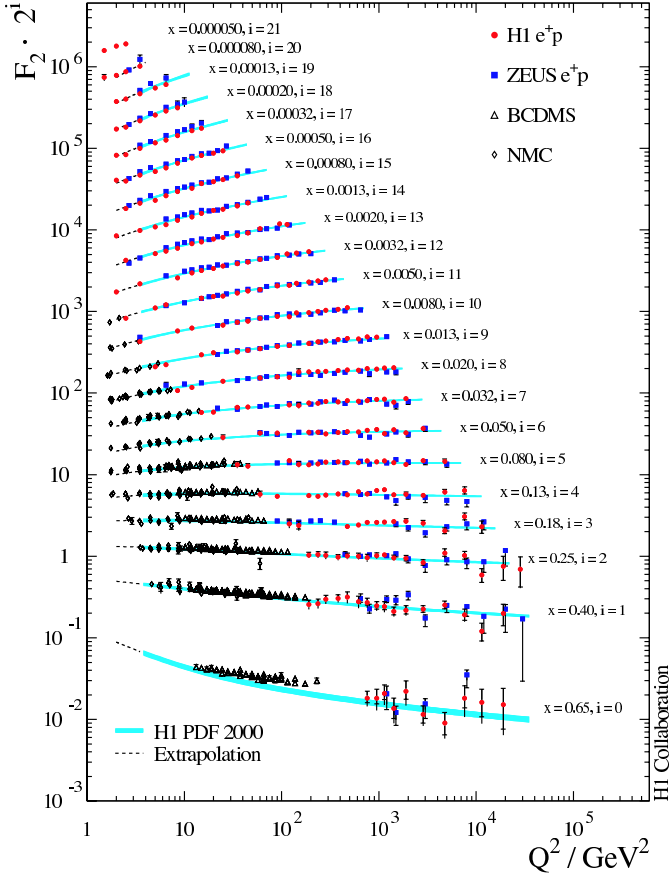}
  \end{minipage}
  \hspace{.03\textwidth}
  \begin{minipage}{0.33\textwidth}
  \vspace{5.5cm} \caption{\label{fig=f2all}\it The proton structure function
  $F_\mathit{2}$ shown as a function of $Q^\mathit{2}$ for fixed values of
  $x$. QCD fits and results from fixed target experiments are also shown.}
  \end{minipage}    
  \vspace{-.8cm}
\end{center}  
\end{figure}

The proton structure function $F_{2}=x\sum
e_{q}^{2}(q(x,Q^2)+\bar{q}(x,Q^2))$, defined as the electromagnetic
contribution to ${\cal F}_{2}$, has been precisely measured by H1 and ZEUS
using neutral current data.  The most recent results are shown in
Fig.~\ref{fig=f2all} together with data from fixed target experiments.  The
present precision reached is of the order of $2-3$\%, apart from
measurements at large $Q^2$ or at the edges of the acceptance region.

The data shown in Fig.~\ref{fig=f2all} cover a large region in $x$ and
$Q^{2}$, clearly exhibiting the well known scaling violations of $F_{2}$ via
the dependence on $Q^{2}$.  In the framework of perturbative QCD
these scaling violations are successfully predicted by the DGLAP
equations~\cite{DGLAP}, which describe the $Q^{2}$--evolution of the quark
and gluon densities in the proton.  At small values of $x$ the behaviour of
$F_{2}$ is dominated by quark-antiquark pair-production arising from gluon
splitting and is given by $\partial F_{2}/\partial \ln Q^{2} \sim \alpha_{s}
\cdot xg(x)$.  DGLAP based QCD fits to these data thus allow, apart from a
determination of the quark content in the proton, also an extraction of the
gluon density and the strong coupling constant $\alpha_{s}$.

The most recent H1 QCD fit~\cite{Adloff:2003uh} (H1 PDF 2000 fit), shown in
Fig.~\ref{fig=f2all}, describes the H1 and ZEUS data well over several
orders of magnitude in $x$ and $Q^{2}$.  A similar fit was performed by the
ZEUS collaboration~\cite{Chekanov:2002pv}.  Both fits are based on the same
standard procedure, for which the parton distribution functions (PDF) are
parametrized at a starting scale $Q_{0}^{2}$ and then evolved to higher
$Q^{2}$ according to the next-to-leading order DGLAP
equations~\cite{Furmanski:1980cm}.  The parameters at $Q_{0}^{2}$ are
determined by a fit of the calculated cross section or $F_{2}$ values to the
data.  The QCD analyses of H1 and ZEUS differ mainly by the amount of non-HERA data
used, the handling of systematic errors, the parametrization at
$Q_{0}^{2}$, and the treatment of heavy quarks.

The H1 PDF 2000 fit uses H1 neutral current (NC) and charged current (CC)
data only.  It determines the gluon density $g(x)$ and four up and down
combinations $U=u+c$, $\bar{U}=\bar{u}+\bar{c}$, $D=d+s$ and
$\bar{D}=\bar{d}+\bar{s}$ from which the valence densities $u_{v}-U-\bar{U}$
and $d_{v}=D-\bar{D}$ can be derived.  In contrast, the most recent
ZEUS QCD analysis~\cite{Chekanov:2002pv} includes the ZEUS NC measurements
together with $\mu p$ and $\mu d$ results from BCDMS, NMC and E665, and CCFR
$\nu {\it Fe}$--scattering data.

\begin{figure}[tbp]
 \begin{center}
   \begin{minipage}{0.4304\textwidth} 
   \includegraphics[width=\textwidth]{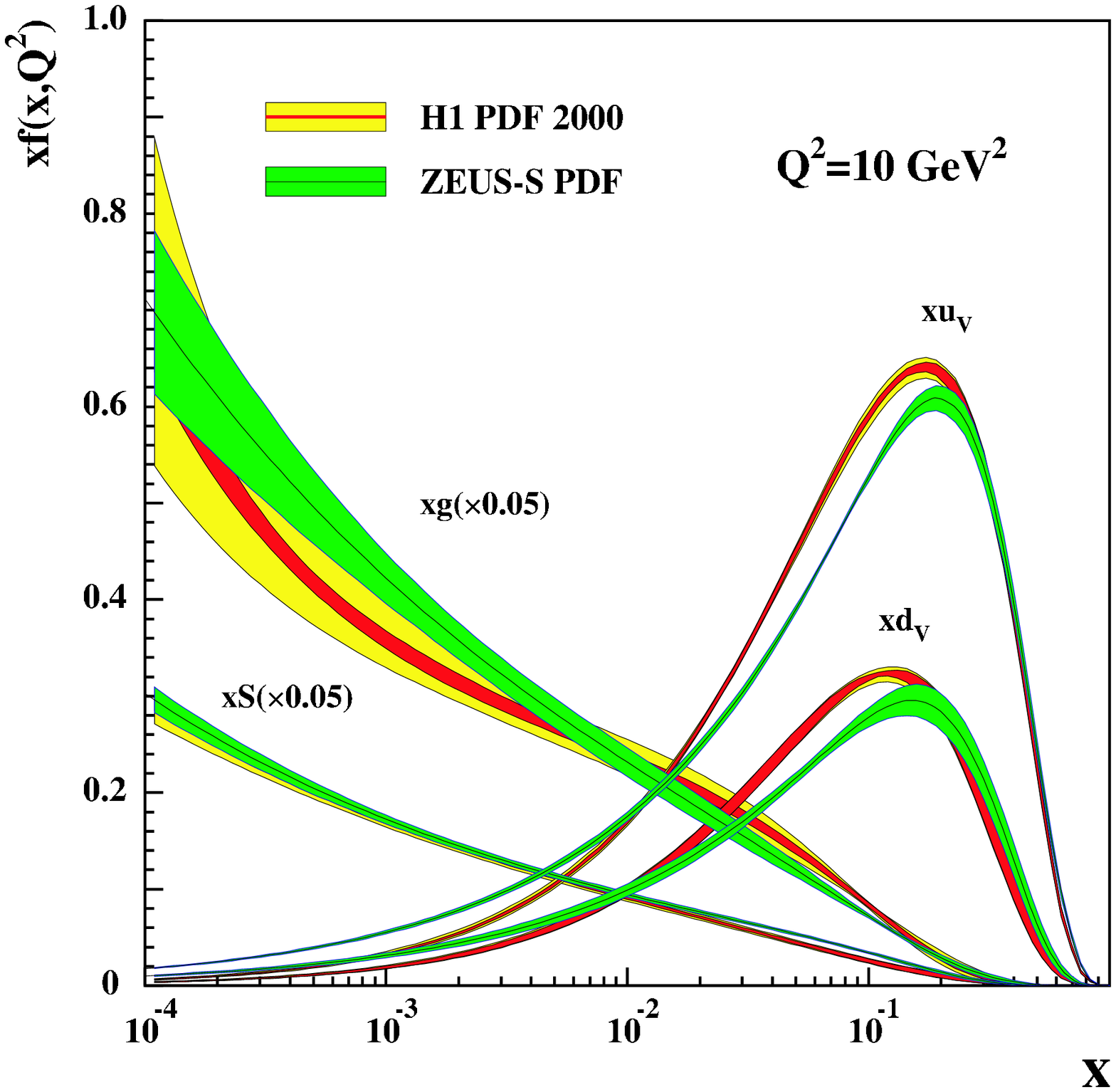}
   \end{minipage}\hspace{0.05\textwidth}
   \begin{minipage}{0.4098\textwidth}  
   \includegraphics[width=\textwidth]{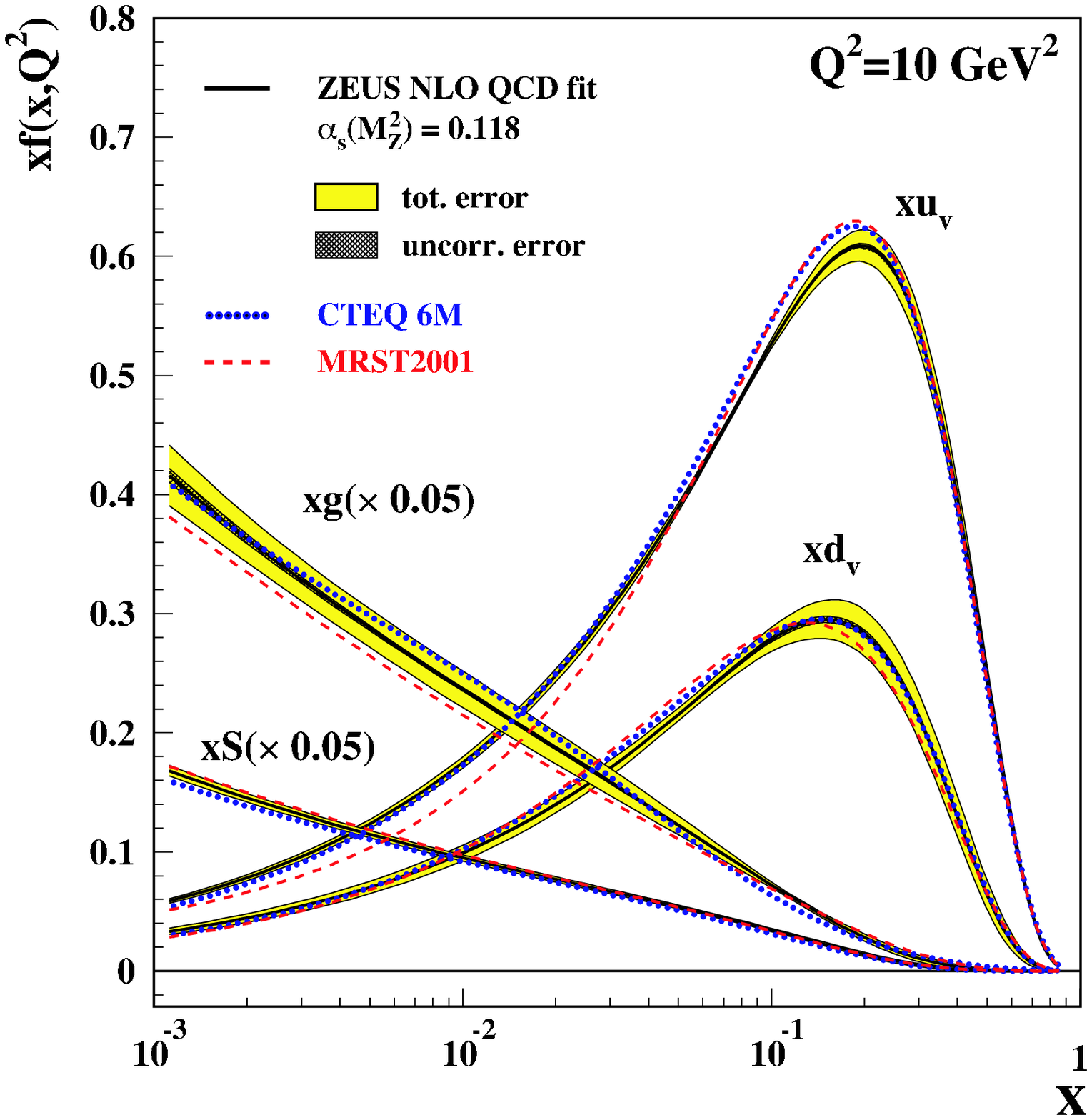}
   \end{minipage} \hspace{0.03\textwidth}
   \caption{\label{fig=PDFvergl}\it (a)~Comparison of PDFs obtained from the
   H1 PDF 2000 and the ZEUS NLO QCD fit.  Shown are the valence quark
   distributions $xu_{v}$ and $xd_{v}$, together with the gluon ($xg$) and
   sea ($xS$) quark densities both scaled down by a factor of 20.
   (b)~Comparison of the ZEUS NLO QCD fit with the global analyses
   MRST2001~\cite{Martin:2001es} and CTEQ6~\cite{Pumplin:2002vw}.}
   \vspace{-.5cm}
 \end{center}  
\end{figure}

The results of these fits at $Q^{2}=10$\,GeV$^{2}$ are compared in
Fig.~\ref{fig=PDFvergl}a, which shows the valence quark distributions
$xu_{v}$ and $xd_{v}$, together with the gluon ($xg$) and sea ($xS$) quark
densities both scaled down by a factor of 20.  The two results are
consistent at the $5$ to $10$\% level and also agree with the results from
global analyses as can be seen from the comparison in
Fig.~\ref{fig=PDFvergl}b.  This is remarkable in view of the different
methods and the different data sets used.  Compared to the ZEUS results the
systematic uncertainty of the H1 PDFs are, however, clearly larger.  This is
expected as in H1 PDF 2000 fit only H1 data are used, which have limited
sensitivity to the valence quark distributions.  When including the BCDMS
data the uncertainties on the H1 result are substantially reduced.

In the central H1 and ZEUS fits, $\alpha_s$ is kept fixed.  If treated as a
free parameter ZEUS obtains $\alpha_s(M_{\rm Z}^2)=0.1166\pm
0.0052$~\cite{Chekanov:2002pv}, with an additional renormalization scale
uncertainty of $\pm0.004$.  This is in agreement with an earlier de\-di\-ca\-ted
H1 analysis yielding $\alpha_s(M_{\rm Z}^2)=0.1166^{+0.0019}_{-0.0018} \pm
0.005$~\cite{Adloff:2000qk}, where the second error contribution is again
estimated by varying the renormalization scale.  This latter uncertainty is
expected to be considerably reduced by full NNLO calculations expected to be
completed soon~\cite{Moch:2002sn}.

\subsection{The Longitudinal Structure Function $F_{L}$}

In the one-photon exchange approximation, which is applicable for $Q^2$-values up
to $Q^2 \approx 1000$~GeV$^2$, the deep-inelastic scattering cross section 
given in equation~(\ref{eqn=NCgen}) reduces to
\begin{equation}
     \frac{\d^2\sigma}{\d x\,\d Q^2} =  
	   \frac{2\pi\alpha^2}{xQ^4} 
	    \left( Y_+\,F_2(x,Q^2) - y^{2}F_L(x,Q^2) \right) \;\; ,
\end{equation}
with $F_{2}$ and $F_{L}$ representing the electromagnetic
($\gamma$--exchange only) contribution to ${\cal F}_{2}$ and ${\cal
F}_{L}$.  An extraction of the longitudinal structure function
$F_{L}$ is therefore only possible at high values of $y$, where the
$F_{L}$--contribution becomes significant.  In order to
disentangle the contributions from $F_{2}$ and $F_{L}$ in a model independent
manner, measurements
of $\d^2\sigma/\d x\,\d Q^2$ at fixed $x$ and $Q^{2}$ for different
values of $y$ are necessary.  A direct measurement of $F_{L}$ thus
requires $ep$--scattering data at different center-of-mass energies,
which up to now are not available at HERA\footnote{The difference in
the center-of-mass energy due to the change in the proton beam energy
from $820$ to $920$\,GeV is not sufficient for a measurement of
$F_{L}$ at HERA.}.

\begin{figure}[tbp]
 \begin{center}
   \begin{minipage}{0.578\textwidth}
   \includegraphics[width=\textwidth]{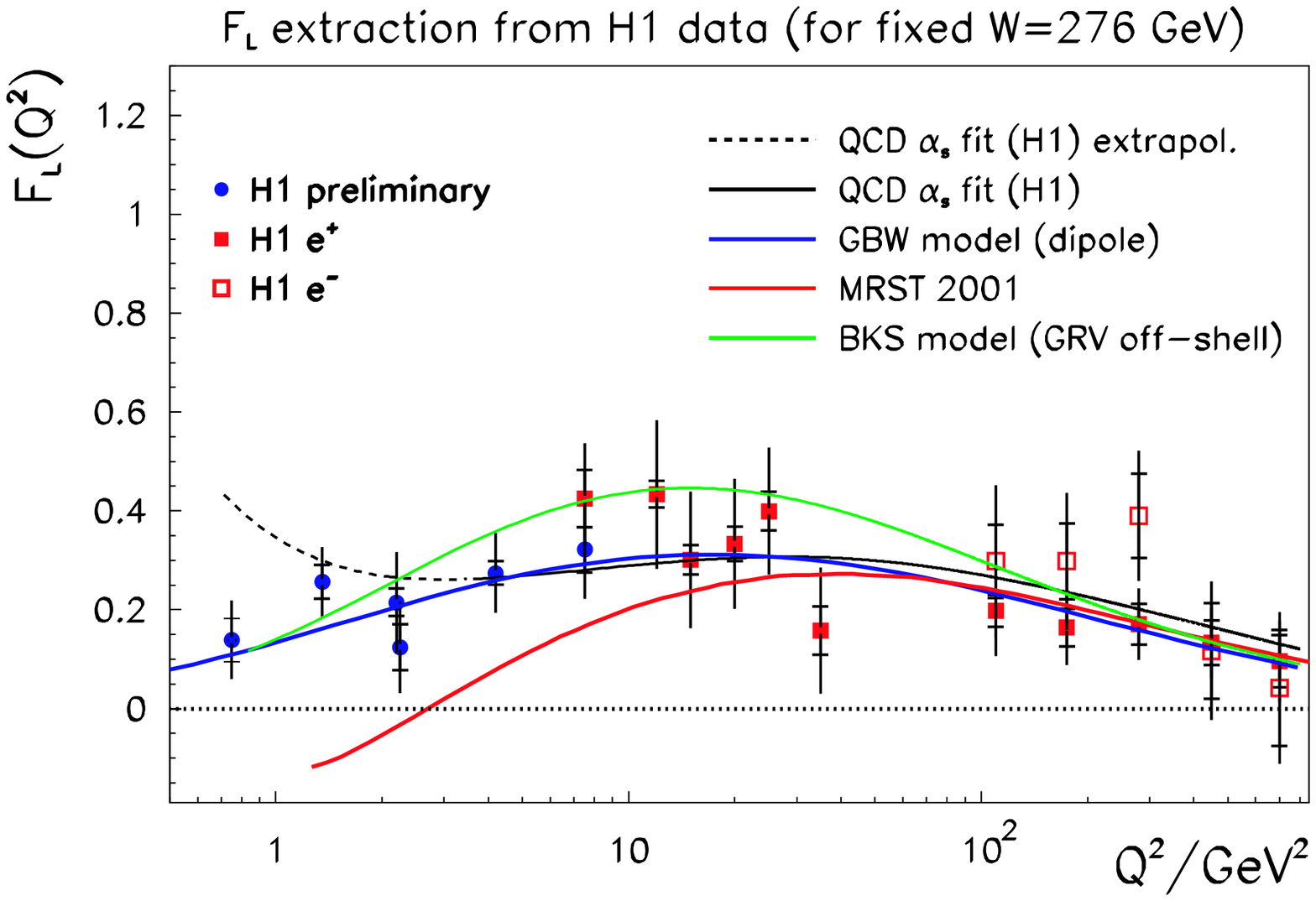}
   \end{minipage} 
   \begin{minipage}{0.402\textwidth} 
   \vspace{.35cm}
   \includegraphics[width=\textwidth]{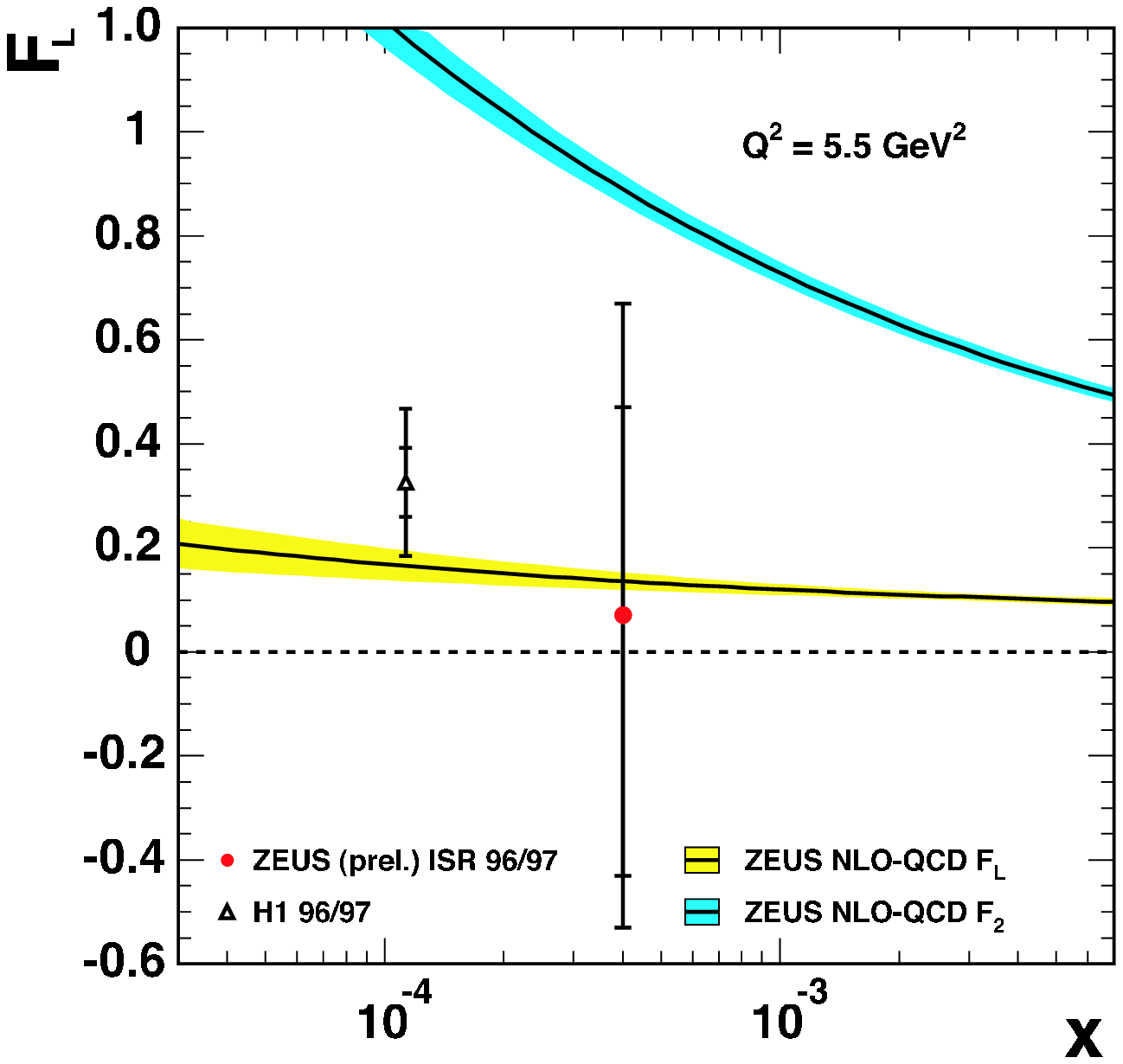}
   \end{minipage}    
   \caption{\label{fig=fl}\it (a) The $Q^\mathit{2}$ dependence of the
   longitudinal structure function $F_\mathit{L}$ at fixed $y \approx \mathit{0.75}$
   ($\,W= \mathit{276}$\,GeV~$\approx \sqrt{ys}$), summarizing the $F_\mathit{L}$ values
   extracted from H1 cross section
   measurements~\cite{Adloff:2000qk,H1_EPS01_lowQ2, H1_EPS03_FL}.  (b)
   Direct $F_\mathit{L}$--measurement at $x=\mathit{4}\cdot\mathit{10^{-4}}$
   and $Q^\mathit{2}=\mathit{5.5}$\,GeV$^\mathit{\,2}$ extracted from ZEUS ISR
   data~\cite{ZEUS_EPS03_FL}.} \vspace{-.5cm}
 \end{center}  
\end{figure}

An indirect determination of $F_{L}$ is, however, possible using DIS
cross section measurements at high values of $y$.  Several extraction
methods have been developed by the H1 collaboration~\cite{Adloff:2000qk,%
H1_EPS03_FL} all of which make use of the different behaviour of the reduced
measured cross section $\sigma_{r}=\frac{\d^2\sigma}{\d x\,\d Q^2}
\frac{xQ^4}{2\pi\alpha^2 Y_{+}} = F_{2} -\frac{y^{2}}{Y_{+}} F_{L}$, 
and the structure function $F_{2}$, which is extrapolated to high
$y$ using low--$y$ data.  Fig.~\ref{fig=fl}a shows the $Q^{2}$ dependence of
the longitudinal structure function $F_{L}$ at fixed $y \approx 0.75$ and summarizes
all $F_{L}$ values extracted by H1 using these methods.  The data points
are in good agreement with most of the theoretical
predictions~\cite{Golec-Biernat:1998js,Badelek:1996ap} shown; at low $Q^{2}$
the MRST 2001 parametrization~\cite{Martin:2001es} is disfavoured.
Concerning the H1 QCD fit note that it is quoted not to be applicable in the
region below $Q^{2}=1$~GeV$^{2}$~\cite{Adloff:2000qk}.
 
ZEUS has recently presented the first direct HERA measurement of $F_{L}$
using initial state radiative (ISR) events~\cite{ZEUS_EPS03_FL}.  For these
events the center-of-mass energy of the $ep$--reaction is reduced due to
emission of a hard photon off the initial state electron.  The result is
shown in Fig.~\ref{fig=fl}b.  Although the measurement is not very precise, it
is clearly consistent with the expectations from perturbative QCD.

\subsection{The Measurement of $xF_{3}$}

\begin{figure}[tbp]
 \begin{center}
   \begin{minipage}{0.495\textwidth} 
   \vspace{0.70cm}                   
   \includegraphics[width=\textwidth]{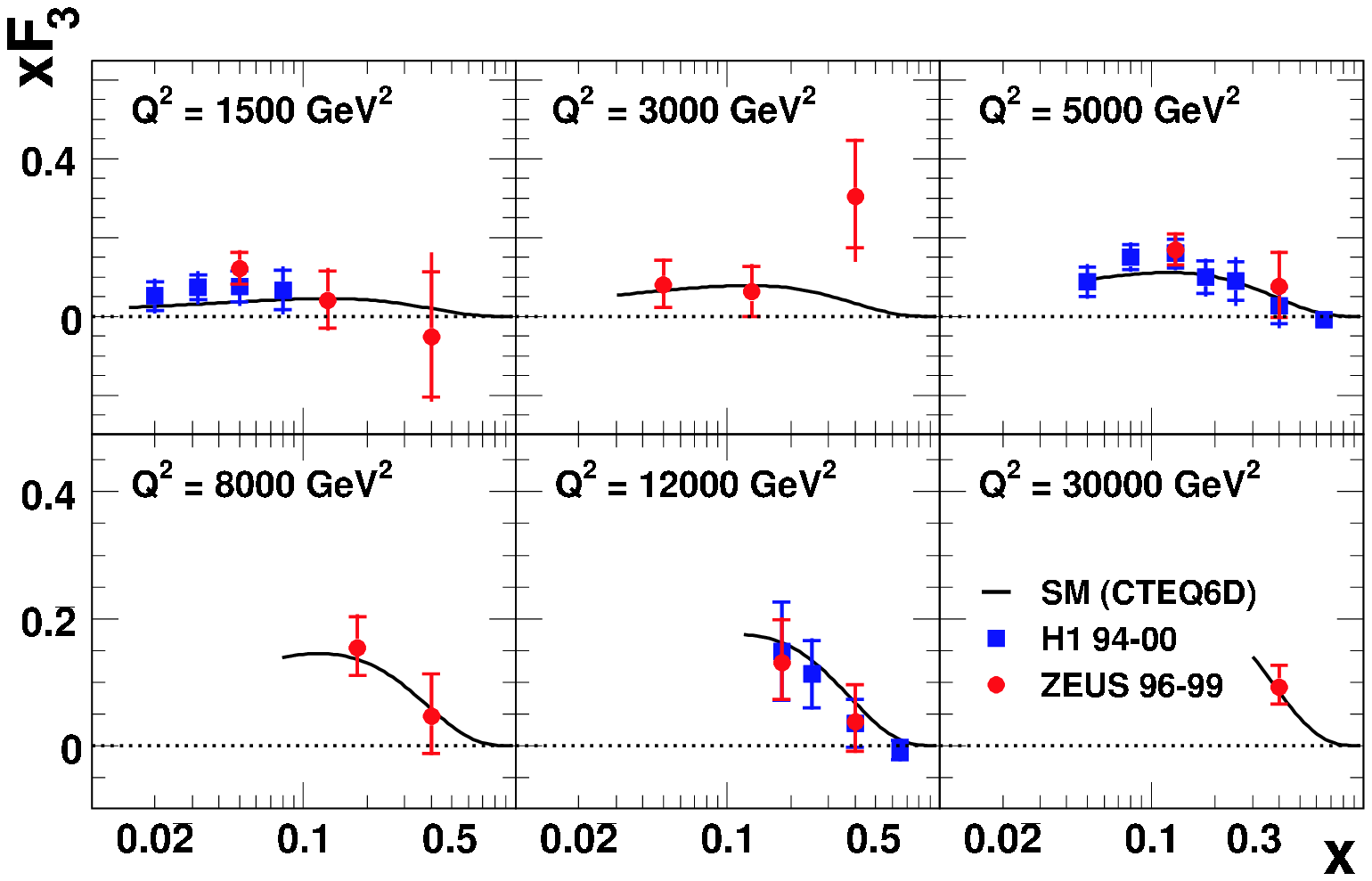}
   \end{minipage} \hspace{0.03\textwidth} 
   \begin{minipage}{0.378\textwidth} 
   \includegraphics[width=\textwidth]{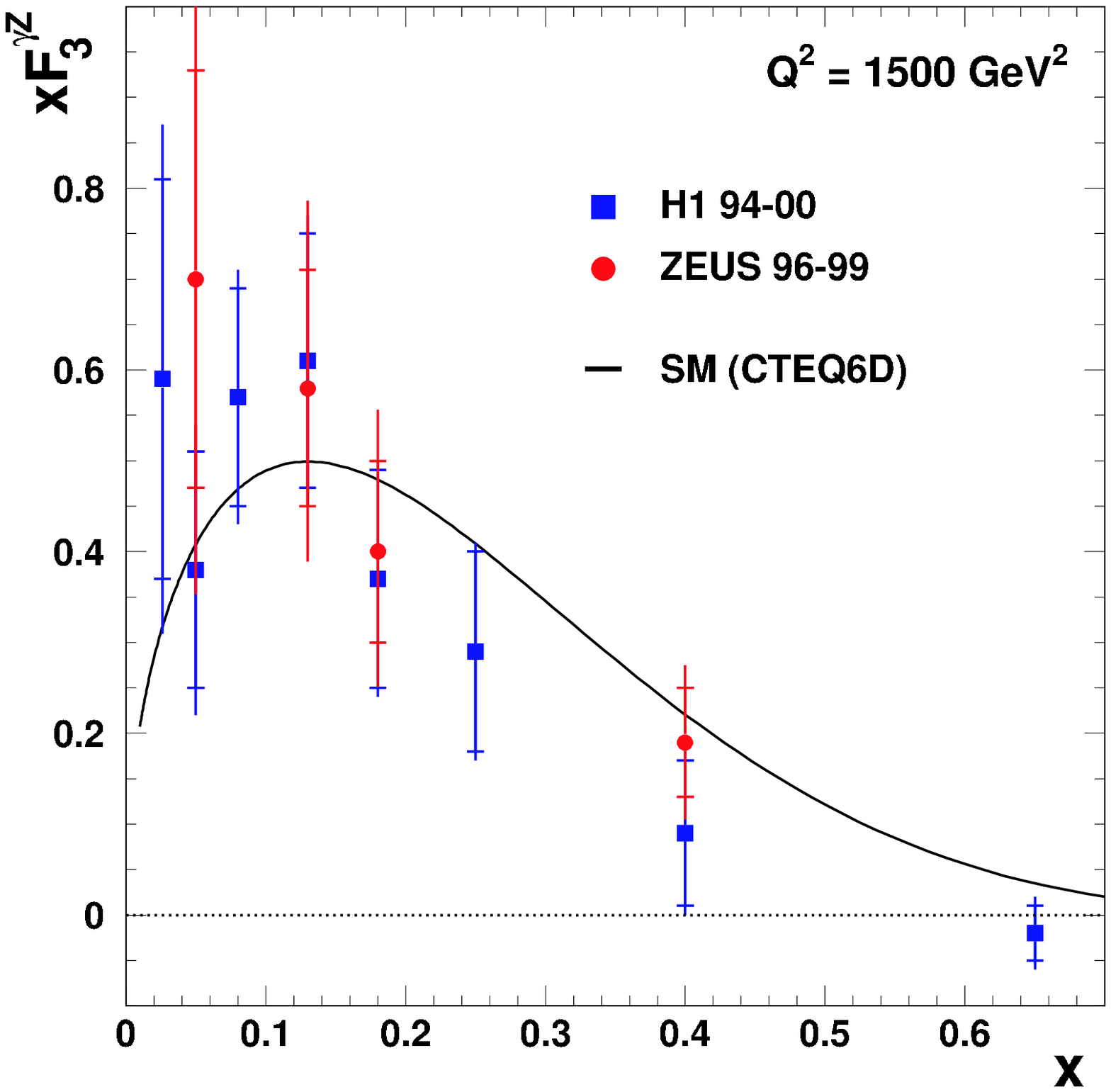}
   \end{minipage} \hspace{0.01\textwidth} 
	\vspace{-.0cm} \caption{\label{fig=xF3}\it (a) The generalized
	structure function $x{\cal F}_{3}$, as extracted by H1 and ZEUS. The
	data are plotted at fixed $Q^{\mathit{2}}$ as a function of $x$.  (b)
	Structure function $xF_{3}^{\gamma\Z}$ as a function of $x$ at
	$Q^{2}$=1500~GeV$^\mathit{\,2}$.  All data are compared to predictions based on
	the CTEQ6D parametrization~\protect\cite{Pumplin:2002vw} of the
	proton PDFs.  Inner error bars represent the statistical, outer
	error bars the total error.  } \vspace{-.5cm}
 \end{center}  
\end{figure}

The difference between the $\e^{-}\p$ and the $\e^{+}\p$ NC cross sections can be
used to extract the generalized structure function $x{\cal F}_{3}$ and ---
since at HERA the dominant contribution to $x{\cal F}_{3}$ comes from the
$\gamma\Z$--interference --- to evaluate the structure function
$xF_{3}^{\gamma\Z}$, which is more closely related to the quark structure of
the proton.  Fig.~\ref{fig=xF3}a shows the $x{\cal F}_{3}$ $x$-dependence
as measured by the H1 and ZEUS
experiments~\cite{Adloff:2000qj,Chekanov:2002ej} in six different bins of
$Q^{2}$. Here, $x{\cal F}_{3}$ is obtained from\footnote{If data at different
center-of-mass energies are used the expression has to be modified slightly.}
\begin{equation}
     x{\cal F}_3 = \frac{1}{2Y_-} \left[ \tilde{\sigma}^{-}_{\rm NC} - \tilde{\sigma}^{+}_{\rm NC}  \right] 
\end{equation}
where the contribution of ${\cal F}_{L}$ can again be neglected; $\tilde{\sigma}^{-}_{\rm NC}$
and $\tilde{\sigma}^{+}_{\rm NC}$ represent the reduced neutral current cross sections
obtained from (\ref{eqn=NCgen}) by removing the trivial $2\pi\alpha^2$/$xQ^4$ dependence.
The results are in good agreement with the QCD prediction.

From these data H1 and ZEUS extract~\cite{Adloff:2000qj,Chekanov:2002ej} the
structure function $xF_{3}^{\gamma\Z} = 2 e_{q} a_{q}
\left[q-\bar{q}\right]$ dividing $x{\cal F}_{3}$ by the factor
$-a_{e}\kappa_{w}Q^{2}/(Q^{2} + M_{\Z}^{2})$; remaining contributions
arising from pure $\Z$-exchange are estimated to be less than 3\% and hence
neglected.  Fig.~\ref{fig=xF3}b shows $xF_{3}^{\gamma\Z}$ as a function of
$x$ for $Q^{2}$=1500~GeV$^2$. 

The presented measurement yields first direct information on the valence quark
content of the proton at high $Q^{2}$.  It is consistent with zero at large
$x$ rising to a maximum at $x \approx 0.1$.  To quantify the level of
agreement between data and theory the following sum rule has been
formulated~\cite{Rizvi:2000qf} in analogy to the Gross Llewellyn-Smith sum
rule~\cite{Gross:1969jf}: $\int_{0}^{1} F_{3}^{\gamma\Z} \d x \approx 5/3$.
Using the H1 data integration yields~$\int_{.02}^{.65} = 1.28\pm 0.20$.  The
corresponding integral obtained for the H1 PDF fit gives $1.06\pm0.02$
in agreement with the measured value.

\section{Jet Physics}
\label{jets}

Jet production in electron-proton collisions at HERA provides a unique
testing ground for Quantum Chromodynamics~(QCD).  Apart from the
determination of the strong coupling constant $\alpha_s$, $ep$~jet data at large
transverse momenta, $E_T$, may
especially be used to gain insight into the dynamics of the exchanged parton
cascade, whose structure is probed by the high-$E_T$ dijet system.

\subsection{$\alpha_s$ Determination}

Jet production at large $Q^2$ and large $E_T$ has been intensively studied
by both the H1 and the ZEUS collaboration.  In this region pertrubative QCD
holds such that it is possible to determine the strong coupling constant
$\alpha_s$.  The renormalization
scale $\mu_r^2$ at which $\alpha_s$ is determined is given, depending on the
process studied, by either $Q^2$ or $E_T^2$.

\begin{figure}[tbp]
 \begin{center}
   \begin{minipage}{0.534\textwidth}
   \vspace{-0.15cm}   
   \includegraphics[width=\textwidth]{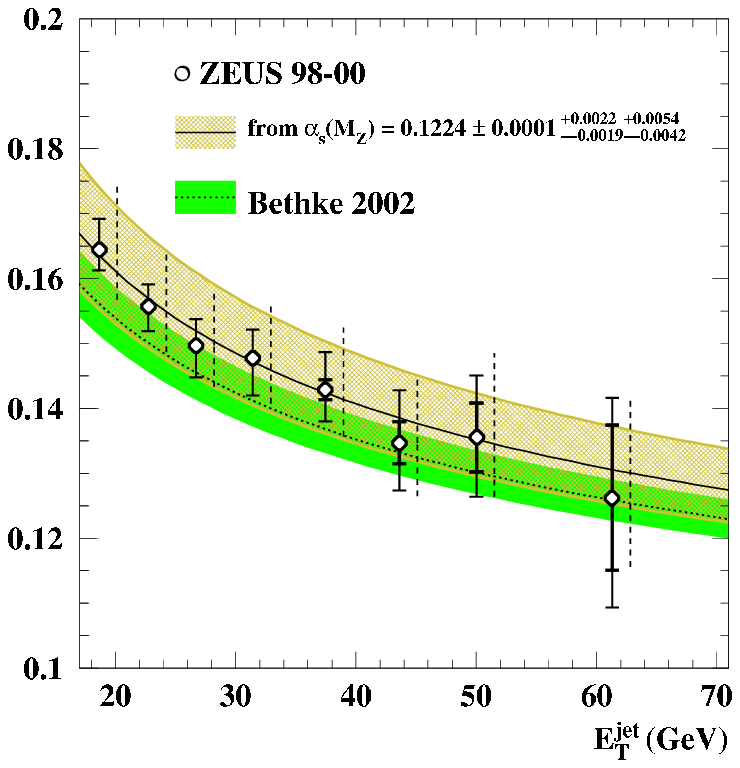}
   \end{minipage}    
   \begin{minipage}{0.434\textwidth} 
   \includegraphics[width=\textwidth]{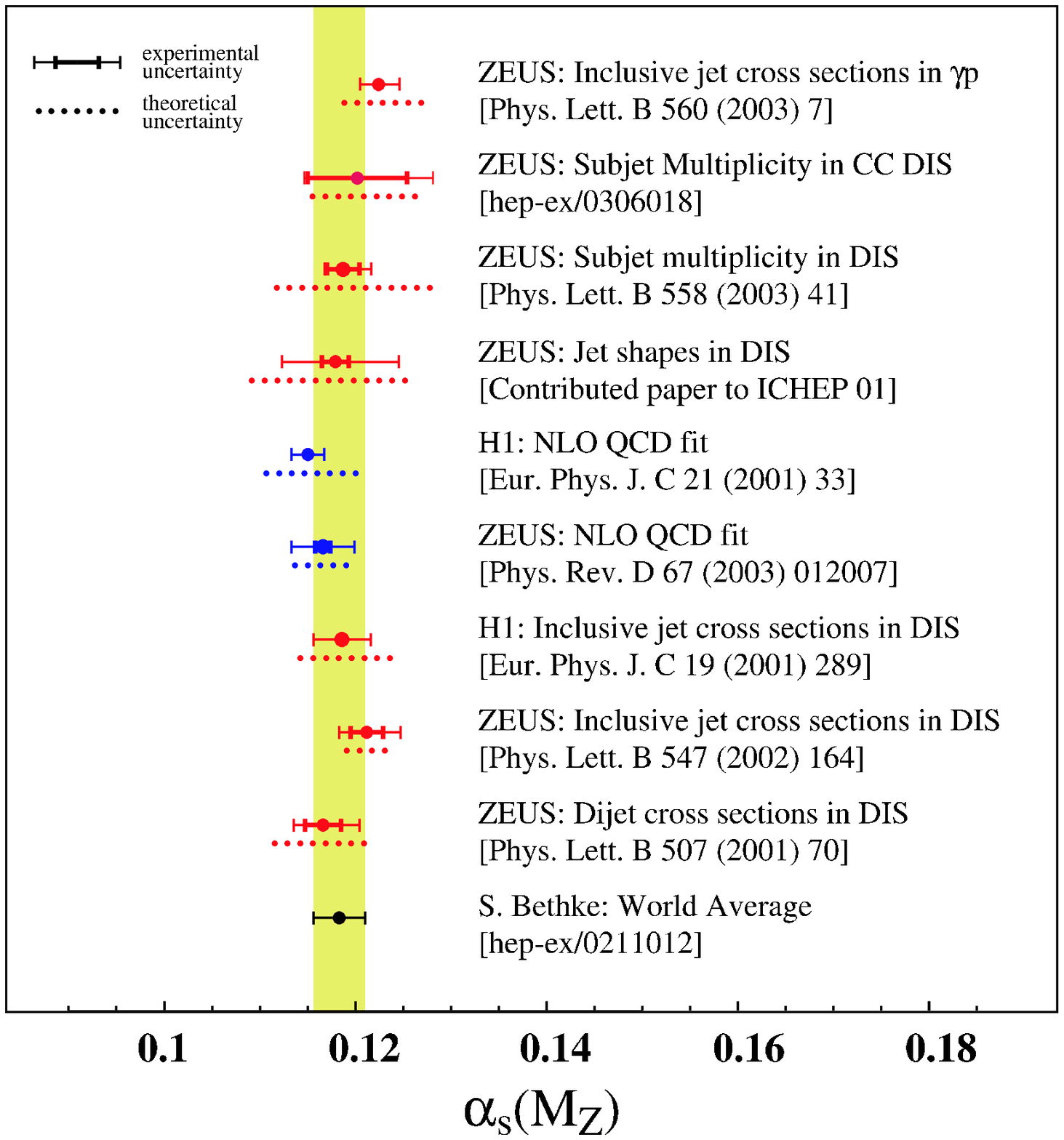}
   \end{minipage} 
   \caption{\label{fig=alphas}\it (a) The $\alpha_s(E_T)$ values determined
   from a QCD fit to the differential $\gamma p$ cross section $d\sigma$/$d
   E_T$ in different $E_T$ regions (open circles).  The solid line
   represents the prediction of the renormalization group equation obtained
   from the $\alpha_s(M_Z)$ central value as determined in the analysis; the
   light-shaded area displays its uncertainty.  (b) Summary of HERA
   $\alpha_s$ measurements in comparison to world
   average~\cite{Bethke:2002rv}.}  \vspace{-.5cm}
 \end{center}  
\end{figure}

Fig~\ref{fig=alphas}a shows one of the most recent $\alpha_s$
measurements~\cite{Chekanov:2002ru} determined from jet production in
$\gamma p$ interactions ($Q^2 \approx 0$); $\alpha_s$ is shown as a function
of $\mu_r=E_T$ clearly revealing the expected scale dependence of the strong
coupling.  When evolved to the mass of the $Z$--boson the analysis yields
$$\alpha_s(M_Z) = 0.1224 \pm 0.0001\; ({\rm stat.}) ^{+0.0022}_{-0.0042}\;
({\rm exp.})  ^{+0.0054}_{-0.0042}\; ({\rm theo.})$$
in agreement with the world average~\cite{Bethke:2002rv} and other
$\alpha_s$-measurements performed at HERA, summarized in Fig.~\ref{fig=alphas}b.
Concerning the experimental uncertainty this value is presently the most
precise measurement of $\alpha_s$ from jets in $ep$--scattering.

\subsection{Probing Parton Dynamics}

HERA jet data cover a large range of $Q^{2}$, Bjorken-$x$ 
and the transverse energy, $E_{T}$, of the observed jets.  
At low~$x$, HERA dijet data may be used to gain insight into
the dynamics of the parton cascade typically exchanged in low-$x$
lepton-proton interactions.

Special insight into small-$x$ dynamics can be gained from inclusive dijet data by
studying the behavior of events with a small azimuthal separation,
$\Delta\phi^{\ast}$, between the two hardest jets as measured in the
hadronic center-of-mass system~\cite{SZCU,FORSetc}.  Partons entering the
hard scattering process with negligible transverse momentum, $k_t$, as
assumed in the DGLAP formalism~\cite{DGLAP}, lead at leading order to a
back-to-back configuration of the two outgoing jets with
$\Delta\phi^{\ast} \sim 180^{\circ}$.  Azimuthal jet separations different from
$180^{\circ}$ occur due to higher order QCD effects.  However, in models which
predict a significant proportion of partons entering the hard process with
large $k_t$, the number of events with small $\Delta\phi^{\ast}$ should
increase.  This is the case for the BFKL~\cite{BFKL} and CCFM~\cite{CCFM}
evolution schemes.

\begin{figure}[tbp]
 \begin{center}
   \begin{minipage}{0.47\textwidth} 
   \includegraphics[width=\textwidth]{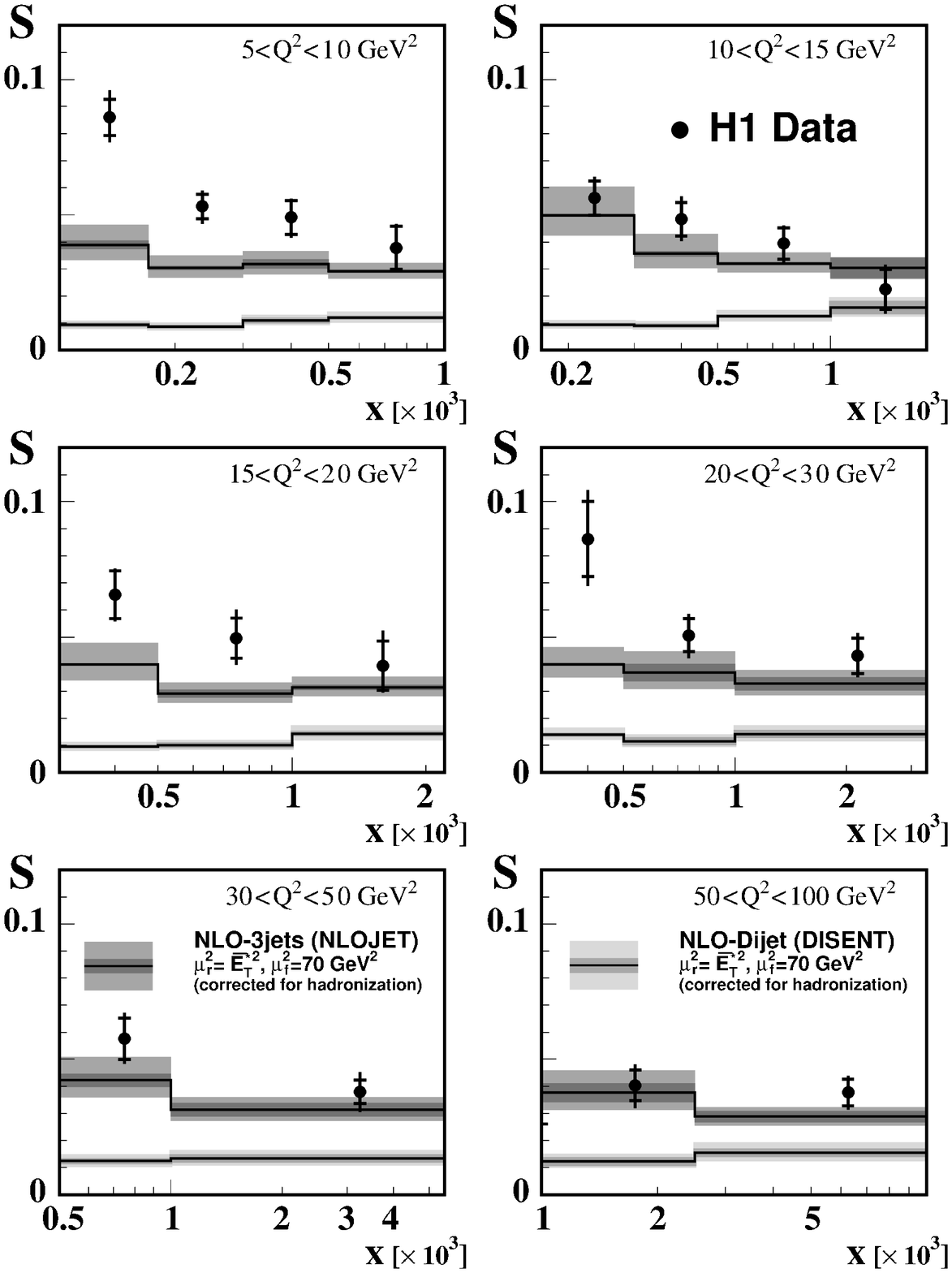}
   \end{minipage} \hspace{0.02\textwidth}
   \begin{minipage}{0.47\textwidth} 
   \includegraphics[width=\textwidth]{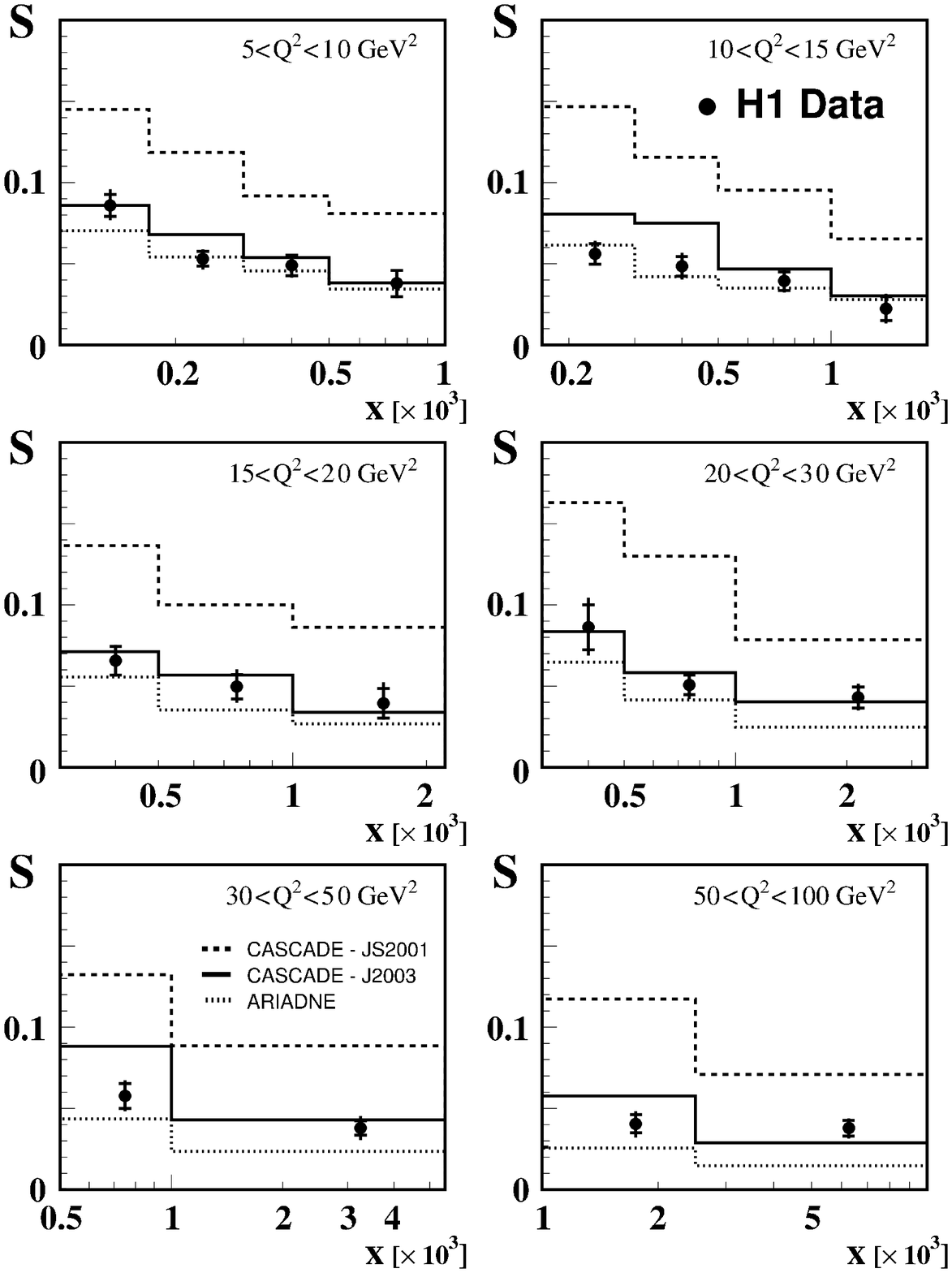}
   \end{minipage} 
   \caption{\label{fig=SDistr}\it Ratio $S$ of events with small azimuthal
   separation \mbox{($\Delta\phi^{\ast}$$<$$\mathit{120}^{\circ}$)} of the
   two most energetic jets with respect to the total number of inclusive
   dijet events given as a function of Bjorken-$x$ and $Q^\mathit{2}$~\cite{Roman}.
   (a) H1 data in comparison to NLO-dijet and 3-jet calculations.\,(b) H1
   data in comparison to predictions from A\-RIAD\-NE and CASCADE.}
   \vspace{-.5cm}
 \end{center}  
\end{figure}

The H1 collaboration has studied the ratio $S$~\cite{Roman} of the number of
events $N_{\rm dijet}$ with an azimuthal jet separation of
$\Delta\phi^{\ast}< \alpha$ relative to all dijet events as proposed
in~\cite{SZCU}.  Fig~\ref{fig=SDistr} shows the $S$ distribution for
$\alpha=120^{\circ}$ as a function of $x$ and $Q^2$. For the chosen $\alpha$
the measured values of $S$ are of the order of 5\%. NLO dijet QCD
calculations~\cite{Disent} predict much too low $S$-values.  The additional
hard emission, provided by the NLO 3-jet calculation~\cite{NLOJET}
considerably improves the description of the data, but is insufficient at
low $x$ and low $Q^2$.  A similar description of the data is provided by
RAPGAP~\cite{Rapgap} a DGLAP-based QCD model, which matches LO matrix
elements for direct and resolved processes to $k_{t}$-ordered parton
cascades (not shown).  A good description of the measured ratio $S$ is given
by the ARIADNE program~\cite{Ariadne}, which generates non-$k_{t}$-ordered
parton cascades using the color dipole model~\cite{CDM}.  Predictions based
on the CCFM evolution equations and $k_t$ factorized unintegrated gluon
densities are provided by the CASCADE Monte Carlo program~\cite{Cascade}.
Large differences are found between the predictions for two different
choices of the unintegrated gluon density, both of them describing the H1
structure function, and one of them giving a good description of $S$.  This
measurement thus provides a significant constraint on the unintegrated gluon
density.

\section{Search for Exotic Final States with Leptons}
\label{searches}

\begin{figure}[tbp]
 \begin{center}
   \begin{minipage}{0.537\textwidth}
   \includegraphics[width=\textwidth]{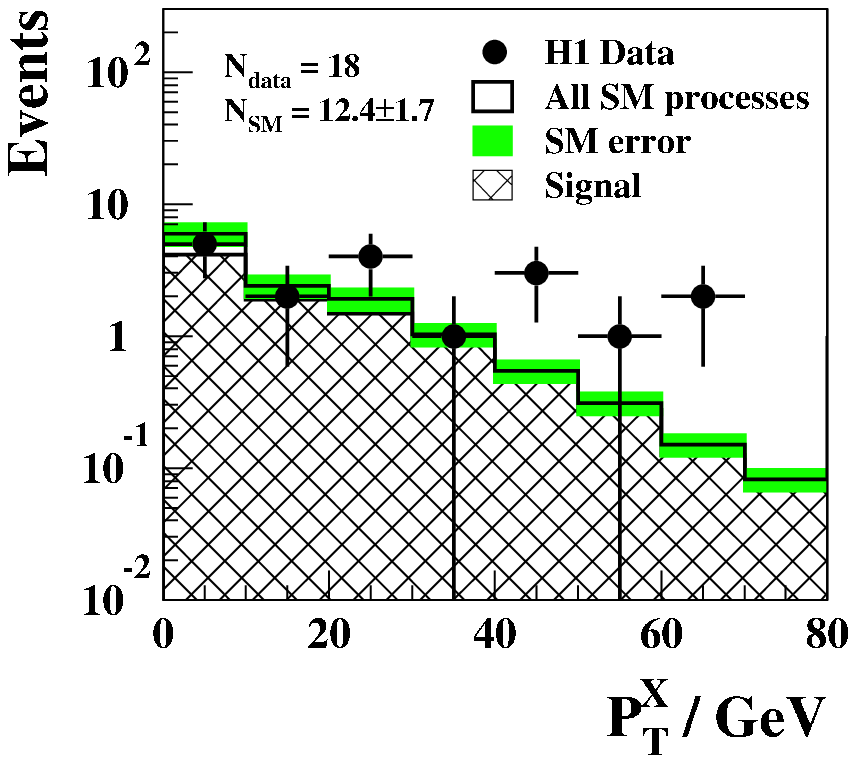}
   \end{minipage}    
   \begin{minipage}{0.443\textwidth} 
   \vspace{-0.80cm}   
   \includegraphics[width=\textwidth]{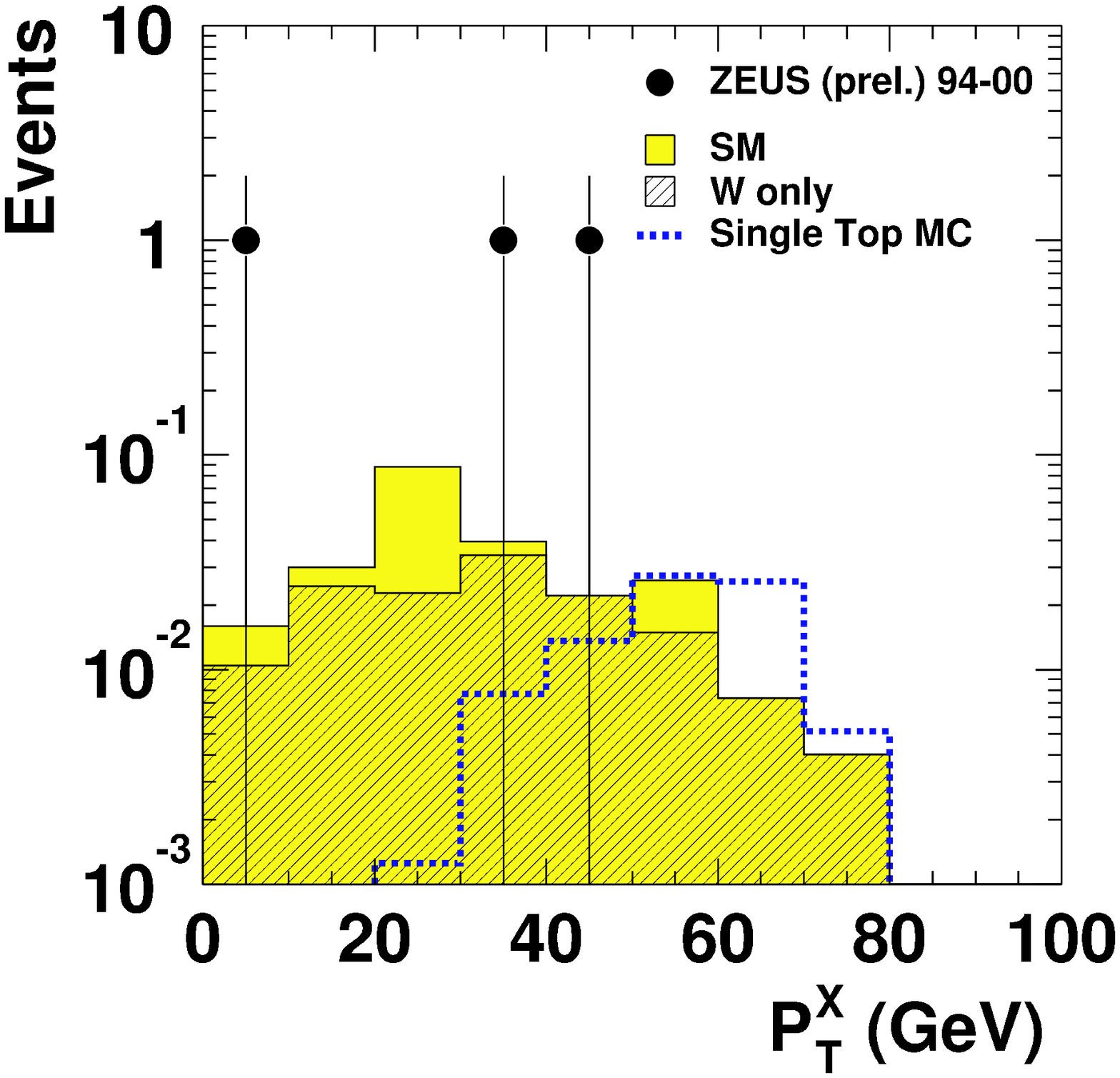}
   \end{minipage} 
   \caption{\label{fig=isol}\it Event distribution for events with missing
   transverse momentum and an isolated electron/muon (a) or and isolated tau
   (b) as a function of $P_T^X$, the transverse momentum of the recoiling
   hadronic system.} \vspace{-.4cm}
 \end{center}  
\end{figure}

By investigating very hard processes in $ep$--scattering H1 and ZEUS have
searched for phenomena beyond the Standard Model (${\cal SM}$) resulting in
a variety of constraints on models involving leptoquarks,
supersymmetry, excited fermions, large extra dimensions etc.  Deviations
from the ${\cal SM}$-expectation have been found for two very distinct
signatures: high energy isolated leptons in events with missing transverse
momentum~\cite{Andreev:2003pm,Chekanov:2003yt, ZEUS_ICHEP02_highpt} and
multi-electron events~\cite{Aktas:2003jg}.  Concerning the first signature,
the H1 collaboration has reported~\cite{Andreev:2003pm} an excess in the
electron and muon channels.  For $P_T^X$$>$40~GeV, where $P_T^X$ is the
transverse momentum of the recoiling hadronic system, H1 observes $3$~($3$)
events in the electron (muon) channel with $0.54\pm 0.11$ ($0.55\pm0.12$)
expected from the ${\cal SM}$ mainly due to production of $W$-Bosons.  No
such events are observed in a similar analysis by ZEUS with $0.94^{+0.11}_{-0.10}$
($0.95^{+0.11}_{-0.10}$) expected~\cite{Chekanov:2003yt}.  However, as
preliminary result~\cite{ZEUS_ICHEP02_highpt} ZEUS finds two events in 
the tau channel for $P_T^X$$>$25~GeV, where only $0.12\pm0.02$ are predicted
by the ${\cal SM}$.  Fig.~\ref{fig=isol} shows the corresponding event
distribution for electrons and muons (taus) observed by H1 (ZEUS) as a function
of $P_T^X$.

\begin{figure}[tb]
\begin{center}
  \begin{minipage}{0.5\textwidth}
  \includegraphics[width=\textwidth]{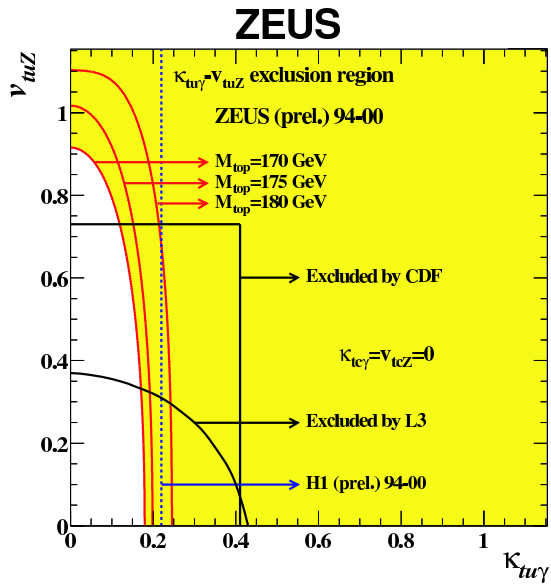}
  \end{minipage}
  \hspace{.03\textwidth}
  \begin{minipage}{0.43\textwidth}
	\vspace{3cm} \caption{\label{fig=anotop}\it ZEUS exclusion regions
	at 95\% CL in the $\kappa_{\mathit tu\gamma}$-$v_{\mathit tuZ}$
	plane for three values of $M_{top}$ assuming $\kappa_{\mathit
	tc\gamma} = v_{\mathit tcZ} = 0$.  Exclusion limits from H1, CDF and
	L4 are also shown.}
  \end{minipage}    
  \vspace{-.8cm}
\end{center}  
\end{figure}   

One possible explanation for the observed isolated lepton events is
anomalous single-top production, $ep \rightarrow etX$, via flavour changing
neutral currents (FCNC).  The anomalous couplings to the photon and the $Z$--boson, 
$tu\gamma$ and $tuZ$, are pa\-ra\-me\-trized by the magnetic coupling
$\kappa_{tu\gamma}$ and the vector coupling $v_{tuZ}$.  Both collaborations
have performed dedicated searches for such processes considering both
leptonic and hadronic decays of the produced top. The H1 experiment sees several
candidate events, a subset of the isolated electron and muon events with missing
transverse momentum. Exclusion limits on $\kappa_{tu\gamma}$ and $v_{tuZ}$ from
HERA~\cite{Chekanov:2003yt,H1_EPS03_singletop} as well as
CDF~\cite{Abe:1997fz} and L3~\cite{Achard:2002vv} are summarized in
Fig~\ref{fig=anotop}.
   
Concerning the multi-electron signature, H1 has reported~\cite{Aktas:2003jg}
an excess of three clean di-electron and three clean tri-electron events
with an invariant di-lepton mass, $M_{12}$, of the two electrons with
highest transverse momentum above $100$~GeV; the corresponding ${\cal SM}$
expectations are $0.30\pm0.004$ and $0.23\pm0.04$, respectively.  A similar
and still preliminary analysis performed by ZEUS~\cite{ZEUS_ICHEP02_highpt}
reveals two di-electron and no tri-electron event, with $0.77\pm0.08$ and
$0.37\pm0.04$ expected.
   
\section{Summary}

Some of the most recent results obtained analyzing hard process in $ep$--interactions have 
been presented. From inclusive deep-inelastic scattering (DIS) data insight into 
the dynamic structure of the proton is gained yielding information on the 
proton parton content and the strong coupling constant $\alpha_s$. Measurements 
of $ep$ jet cross sections provide important tests of perturbative QCD; by  
analyzing various aspects of jet data several values of $\alpha_s$ are obtained all
in agreement with each other, the result from inclusive DIS data and the world average.
Some new insight into small-$x$ dynamics is gained by a very recent result, obtained
studying dijet events with small separation in azimuth of the two most energetic
jets. In general, by investigating hard processes in $ep$--interactions many aspects of the
Standard Model can be tested. Any observed deviation between the data and existing 
theoretical models could hint at signs of new physics. Two such deviations have been observed
at HERA when searching for events with a high-$p_{t}$ isolated lepton and missing 
transverse momentum and multi-electron topologies.

\section{Acknowledgements}
This work was supported by the Bundesministerium f\"ur Bildung,
Wissenschaft, Forschung und Technologie, Germany (contract no. 05H11PEA/6).

\end{document}